\documentclass[conference]{IEEEtran}
\usepackage{cite}

\ifCLASSINFOpdf
 \usepackage[pdftex]{graphicx}
\DeclareGraphicsExtensions{.pdf}
\else
\fi
\ifCLASSOPTIONcompsoc
\usepackage[caption=false,font=normalsize,labelfont=sf,textfont=sf]{subfig}
\else
\usepackage[caption=false,font=footnotesize]{subfig}
\fi
\usepackage{url}
\usepackage{multirow}
\usepackage{amsmath}
\usepackage{slashbox}

\begin{document}
%
\title{An Adaptive Tabu Search Algorithm for Market Clearing Problem in Turkish Day-Ahead Market}


 \author{\IEEEauthorblockN{Nermin Elif Kurt}
\IEEEauthorblockA{Optimization Specialist\\
Energy Exchange Istanbul, Turkey\\
Email: nermin.kurt@epias.com.tr}
 \and
\IEEEauthorblockN{H. Bahadir Sahin}
 \IEEEauthorblockA{Optimization Specialist\\
 Energy Exchange Istanbul, Turkey\\
 Email: bahadir.sahin@epias.com.tr}
 \and
 \IEEEauthorblockN{K\"{u}r\c{s}ad Derinkuyu}
 \IEEEauthorblockA{Dept. of Industrial Engineering \\
 TOBB University of Economy\&Technology\\
 Email: kderinkuyu@etu.edu.tr}
}



%

\IEEEoverridecommandlockouts
\IEEEpubid{\makebox[\columnwidth]{978-1-5386-1488-4/18/\$31.00~
\copyright2018
IEEE \hfill} \hspace{\columnsep}\makebox[\columnwidth]{ }}


\maketitle

\begin{abstract}
In this study, we focus on the market clearing problem of Turkish day-ahead electricity market. We propose a mathematical model by extending the variety of bid types for different price regions. The commercial solvers may not find any feasible solution for the proposed problem in some instances within the given time limits. Hence, we design an adaptive tabu search (ATS) algorithm to solve the problem. ATS discretizes continuous search space arising from the flow variables. Our method has adaptive radius and it achieves backtracking by a commercial solver. Then, we compare the performance of ATS with a heuristic decomposition method from the literature by using synthetic data sets. We evaluate the performances of the algorithms with respect to their solution times and surplus differences. ATS performs better in most of the sets.
\end{abstract}


%
\IEEEpeerreviewmaketitle

\section{Introduction}

Liberalization of Turkish electricity markets started with the first electricity market law 4628 in 2001. In 2013, the law was repealed and replaced by law 6446. In accordance with the new law, the Turkish market operator, Energy Exchange Istanbul (EXIST), was founded in 2015. The main activity areas of EXIST include day-ahead electricity, intra-day electricity and wholesale gas markets. Participants in Turkish day-ahead electricity market (DAM) can offer bids by a two-sided blind auctions one day before the actual delivery date. Then, the market operator announces market clearing prices (MCPs) and bid matchings according to the submitted bids. 

Bidding types in Turkish DAM are similar to the ones in European power exchanges (PXs). There are three different offers in Turkish DAM, that are hourly, block, and flexible bids. An hourly bid is represented by multiple price and volume levels for each period. These volume-price pairs in a period form piece-wise linear demand and supply curves. A block bid contains a single price, volume, start period, and a duration (certain number of consecutive periods). It has to be matched in all of the periods or entirely rejected. There might be a link relation between block bids, such that acceptance of a bid (called \textit{child}) depends on the acceptance of another bid (called \textit{mother}). A flexible bid has a price and a volume information that is valid for only one period. The starting period of the flexible bid is determined by the clearing mechanism among the periods within the flexible bid's interval limit. In near future, EXIST plans to extend the features of the block and flexible bids to satisfy the participants’ needs. Possible extensions are varying quantity per period for both block and flexible bids, and allowing the duration of flexible bids to more than one periods\footnote{https://www.epias.com.tr/en/announcements/market/day-ahead/day-ahead-market-phase-ii-new-order-types}. \cite{oneill2007equilibrium} defines block bids as \textit{non-convex bids}. In this study, we extend the \textit{non-convex bid} definition for both flexible and block bids.

Electric power transmission is limited by the capacities of the transmission elements. When the load exceeds the certain limits, the capacity congestion occurs. Such congestion can be managed with various ways. Market splitting in DAMs is one of the congestion management strategies. Here, the system operator divides DAM into different zones and the market clearing model considers transmission constraints between these zones. This model assumes that there is no major congestion inside a zone during the actual delivery of the electricity. The resulting flows do not necessarily satisfy the Kirchhoff's laws. The current Turkish DAM does not leverage market splitting actively; however, it is legalized by the electricity market regulations since 2009. Turkish Electricity Day Ahead Market Optimization Software (TE-DAMOS)\footnote{https://www.epias.com.tr/en/day-ahead-market/matching. The document explains how the problem is solved by EXIST.} used in EXIST is capable of solving the multi-zonal market clearing problems.

Market clearing in Turkish DAM is a surplus maximization problem. It considers the bid execution conditions, transmission limits and pricing constraints on the bids and flows. Each day, EXIST solves the problem within 10 minutes and announces the market clearing prices, accepted bids, and the optimal flows to the market participants. Thus, the problem should be solved so that at least one good, feasible solution is attained regardless of the problem's computational complexity. 

Proposing solution methodologies for the market clearing problems is a prominent research area in the literature. \cite{martin2014strict} considers block and flexible bids in its formulation and proposes decomposition-based exact and heuristic solution approaches. \cite{madani2015computationally} introduces stronger cuts than \cite{martin2014strict} for only block bid. 

In addition to block and flexible bids, there are also different types of bids in European DAMs as minimum income condition (MIC) orders and prezzo unico nazionale (PUN) orders which are used in Spanish-Portugues and Italian PXs, respectively. These bids are covered in \cite{vlachos2011balancing, blanco2014revenue, madani2017mip}.

In this study, we focus on the market clearing problem of EXIST. We expand the problem statement with extended bid types for different price regions and propose a new mathematical model, i.e., mixed integer quadratically constrained quadratic programming problem (MIQCQP). We design an adaptive tabu search (ATS) algorithm to solve the proposed problem. We create synthetic data sets to solve the MIQCQP problem by using ATS and the heuristic method proposed in \cite{martin2014strict}. We evaluate the performance of these two heuristic methods based on the solution times and relative gaps. 

We contribute to the literature by proposing an algorithm that is able to find at least one solution within 10 minutes, which is a critical time limit for finding a feasible solution for Turkish DAM. In addition, compared to the method in \cite{martin2014strict}, our proposed algorithm provides competitive surplus values.

The organization of the paper is as follows. Section II constructs a mathematical model for the market clearing problem in Turkish DAM. Section III proposes an adaptive tabu search algorithm (ATS) to solve the defined problem. The numerical studies and their results are presented in Section IV. We conclude our paper in Section V. 


\section{Market Clearing  Model}
Non-convex electricity market is a widely used concept in the literature where binary decision variables create a non-convex feasible region. \cite{oneill2005efficient, ruiz2012pricing, van2011linear} propose several clearing rules to deal with such non-convexities. Due to these non-convexities, there are accepted bids with negative surplus and rejected bids with positive surplus. This phenomenon is called \textit{price-matching incompatibilities}. To eliminate these incompatibilities, PXs allow either rejected bids with positive surplus or accepted bids with negative surplus. The price-matching compatibility constraints in Turkish DAM force to accept  non-convex bid if it has positive surplus except child block bids. In that case, the decision on the child block bid must follow the same decision of its mother.

In Turkish DAM, there are also price-flow conditions as discussed in \cite{martin2014strict}. If the flow on a line does not equal to one of its limits for a given period, then there is no price differentiation between the zones that the line connects in that period. If the flow is bounded above/below for a period, the price of the source node of the line must be lower/higher than the price of the sink node in that period. 

In this study, we use piece-wise hourly bid curves and non-convex bids in our model. We assume that non-convex bids share the same properties, such that (1) each non-convex bid can be linked to another bid as in block bids, and (2) duration of a non-convex bid is less than or equal to the length of the interval limit. We model available transmission capacity (ATC) constraints and ramping limits on the transmission lines which bound the flow differentiation between two consecutive periods. We also consider price-flow conditions in our model.

 
\subsection{Indices, Sets, Parameters and Decision Variables }
In this section, we provide indices, sets, parameters, and decision variables to construct a model where piece-wise hourly bids, non-convex bids and transmission constraints are handled.  

\begin {table}[h!]
\begin{tabular}{c l}
Indices and Sets & \\
\hline
$t,T $& Index and set of time periods, $T=\{1,2,...,24\}$. \\
$n,N$ & Index and set of zones. \\
$l, L$ & Index and set of transmission lines \\
$s_l,e_l$ & Source and sink zone of line $l$, $l\in L$.\\
$L^s_n, L^e_n$ & Set of transmission lines starting and ending in \\
& zone  $n \in N$, $L= L^e_n \cup L^s_n$.\\
$i,I_{n,t}^s, I_{n,t}^d $& Index and set of supply and\\
& demand segments in node $n$ in period $t$ where \\
& $I_{n,t} = I_{n,t}^s \cup I_{n,t}^d$.\\
$b,B_n$ & Index and set of non-convex bids in zone $n$. \\
$\Lambda_b$  & Set of non-convex bids that $b$ can be accepted\\
& if $b_\lambda \in \Lambda_b$ is accepted. \\
$T_b$ & Set of periods that $b$ can be started\\
\end{tabular}
\end{table}

\begin {table}[h!]
\begin{tabular}{c l}
Parameters &  \\
\hline
$P_{min},P_{max}$ & Minimum and maximum price limits. \\
$P^0_{i,t},P^1_{i,t}$ & Starting and ending price for segment $i$, in period $t$, \\
$Q_{i,t}$ & Volume of segment $i$, $\forall$ $i\in I_{n,t}$, $n \in N$, $t\in T$.  \\
$P_b,Q_{b,\overline{t},t}$ & Price and volume of a non-convex bid $b$ in period $t$ \\
& if the bid is accepted in period $\overline{t}$, $\forall$ $b\in B$, $\forall$ $\overline{t}\in T_b$, $t \in T$. \\
$\overline{\tau}_{l,t}, \underline{\tau}_{l,t}$ & Upper and lower bound on transmission quantity\\
& of line $l$ in period $t$. \\
$\tilde{\tau}_{l,t}$ & Ramping limit of line $l$ in period $t$. \\
\end{tabular}
\end{table}

Hourly bids are defined by a set of volume-price pairs. These pairs represent maximum/minimum price for the volume that a bidder offers to buy/sell. We construct supply and demand curves for each period by aggregating buy and sale volumes coming from the pairs of hourly bids in that period, respectively. While the supply curve is a non-decreasing function of the quantity, the demand curve has non-increasing property. Each area between the pairs is called a \textit{segment}. If the pairs belong to a demand (supply) curve then this segment is called \textit{demand segment} (\textit{supply segment}). For the supply segments, $P^0_{i,t} < P^1_{i,t}$ for all $i\in I_{n,t}^s$, $t\in T$, and for the demand segments, $P^1_{i,t} < P^0_{i,t}$ for all $i\in I_{n,t}^d$, $t\in T$, $n\in N$. Volume of a segment $i$ is the difference of the volumes of the pairs that create the segment. We assume that $Q_{i,t} \leq 0$ for a supply segment $i$, for all $i\in I_{n,t}^s$ and $Q_{i,t} \geq 0$ for a demand segment $i$, for all $i\in I_{n,t}^d$, $t\in T$, $n\in N$. Indices of the segments are ordered in increasing numbers. A supply segment with the smallest index is the one at the minimum price level and a demand segment with the smallest index is the one at the maximum price level.
 

\begin {table}[h!]
\begin{tabular}{c l}
Decision Variables &\\
\hline
$x_{i,t}$ & Accepted fraction of segment $i$, $\forall$ $i\in I_{n,t}$. \\
$f_{l,t}$ & Flow at line $l$,  $\forall$ $l\in L$, $t \in T$. \\
$y_{b,\overline{t}}$ & 1 if non-convex bid $b$ is accepted in period $\overline{t}$, \\
& 0 otherwise, $\forall$ $b\in B_n $, $n\in N$, $\overline{t}\in T_b$.\\
$p_{n,t}$ & Price at zone $n$ for period $t$, $\forall$ $n\in N$, $t \in T$. \\
$\overline{\mu}_{l,t}, \underline{\mu}_{l,t}$ & 0 if the flow on line $l$ does not equal to upper or \\
& lower bound on transmission quantity in period $t$,\\
& shadow price of the capacity congestion o.w. \\
& $\forall$ $l\in L$, $t \in T$.\\
$\overline{\rho}_{l,t}, \underline{\rho}_{l,t}$ & 0 if the flow on line $l$ in period $t$ is not limited\\
&by the ramping limit, or the shadow price of the  \\
&ramping constraint o.w. $\forall$ $l\in L$, $t \in T$.\\
\end{tabular}
\end{table}

For a line $l$, $f_{lt}>0$ represents flow from $s_l$ to $e_l$ and $f_{lt}<0$ shows the reverse flow. Constant $f_{l,0}$ equals to the flow in the last period of the previous day.

\subsection{Mathematical Model}
In this section, we model the market clearing problem in Turkish DAM which we call as \textit{Model-P}. We use the mathematical programming problems with complementary constraints (MPCC) approach to formulate the model which is similar to \cite{martin2014strict}.

(Model-P)
\begin{equation*}
\begin{split}
    \max g = \sum_{n\in N} \sum_{t\in T}  \Big ( \sum_{i\in I_{n,t}} Q_{i,t}P_{i,t}^0 x_{i,t} &+ Q_{i,t}(P_{i,t}^1-P_{i,t}^0)\frac{x_{i,t}^2}{2} \\
    &+ \sum_{b\in B_n,\overline{t} \in T_b} Q_{b,\overline{t},t} P_b y_{b,\overline{t}} \Big )\\
\end{split}
\end{equation*}
\textit{subject to}
\begin {table}[!h]
\begin{tabular}{l r}
   $\displaystyle \sum_{i \in I_{n,t}} Q_{i,t}x_{i,t} + \sum_{b\in B_n,\overline{t}\in T_b} Q_{b,\overline{t},
    t} y_{b,\overline{t}}  + \sum_{l\in L^s_n} f_{l,t}$\\
    $\quad \quad \quad \quad \quad\quad \quad \displaystyle - \sum_{l\in L^e_n} f_{l,t} = 0, \quad \quad  \quad  \quad  \quad \quad \quad  \forall t\in T, n\in N,$ & (1)\\
    $ x_{i,t} \leq 1,  \quad  \quad  \quad  \quad  \quad  \quad \quad  \quad  \quad  \quad  \quad  \quad \quad  \forall i \in I_{n,t},  t\in T , n\in N, $ & (2) \\
    $\displaystyle  \sum_{\overline{t}\in T_b} y_{b,\overline{t}} \leq 1, \quad  \quad  \quad  \quad \quad  \quad  \quad  \quad  \quad  \quad \quad \quad \quad \quad \forall b\in B_n, n\in N, $ & (3)\\ 
    $\displaystyle  \sum_{\overline{t}\in T_b} y_{b,\overline{t}} \leq \sum_{\overline{t}\in T_{\lambda_b}} y_{\lambda_b,\overline{t}}, \quad  \quad  \quad  \quad  \quad \forall \lambda_b \in \Lambda_b, b\in B_n, n\in N, $ & (4)\\
     $ -\tilde{\tau}_{l,t} \leq f_{l,t} -  f_{l,t-1}  \leq \tilde{\tau}_{l,t},  \quad  \quad  \quad \quad \quad \quad \quad  \quad  \quad  \forall l\in L, t\in T, $ & (5) \\
     $\displaystyle (1-\sum_{\overline{t}\in T_b} y_{b,\overline{t}}) \sum_{t\in T} (P_b-p_{n,t})Q_{b,\overline{t},t} \leq 0, \quad$\\
     $\quad\quad\quad\quad\quad\quad\quad\quad \quad\quad\quad\quad\quad\quad\quad  \quad\quad \forall \overline{t} \in T_b, b\in B,  \Lambda_b =\emptyset, $ &  (6)\\
     $\displaystyle  (\sum_{ \overline{t} \in T_{\lambda_b}} y_{\lambda_b,\overline{t}}-\sum_{\overline{t}\in T_b} y_{b,\overline{t}}) \sum_{t\in T} (P_b-p_{n,t})Q_{b,\overline{t},t} \leq 0$, \\
     $ \quad \quad \quad \quad \quad \quad \quad \quad \quad \quad \quad \quad \quad  \forall \overline{t} \in T_b, \lambda_b \in \Lambda_b, b\in B, \Lambda_b \neq \emptyset, $ & (7) \\
    $\displaystyle  ( -\overline{\tau}_{l,t} +f_{l,t}) \overline{\mu}_{l,t} = 0, \quad \quad \quad \quad \quad \quad \quad \quad \quad \quad \quad \forall l\in L, t\in T, $ & (8)  \\
   $ \displaystyle ( \underline{\tau}_{l,t}-f_{l,t})\underline{\mu}_{l,t} = 0, \quad \quad \quad \quad \quad \quad \quad \quad \quad \quad \quad \quad \forall l\in L, t\in T,$  &  (9)  \\
   $\displaystyle  ( -\tilde{\tau}_{l,t} -f_{l,t-1}+f_{l,t}) \overline{\rho}_{l,t} = 0,  \quad \quad \quad \quad \quad \quad \quad \forall l\in L, t\in T, $   & (10)  \\
   $ \displaystyle ( -\tilde{\tau}_{l,t} +f_{l,t-1}-f_{l,t})\underline{\rho}_{l,t} = 0, \quad \quad \quad \quad \quad \quad \quad \forall l\in L, t\in T, $   & (11)  \\
   $\overline{\mu}_{l,t}-\underline{\mu}_{l,t} + \overline{\rho}_{l,t}- \underline{\rho}_{l,t} -\overline{\rho}_{l,t+1}+ \underline{\rho}_{l,t+1} = p_{e_l,t}-p_{s_l,t}, $\\
   $\quad \quad \quad \quad  \quad \quad \quad \quad \quad  \quad \quad \quad \quad \quad  \quad \quad \quad \quad \quad   \forall l\in L, t\in T\setminus 24,$ &(12) \\
    $\overline{\mu}_{l,t}-\underline{\mu}_{l,t} + \overline{\rho}_{l,t}- \underline{\rho}_{l,t} = p_{e_l,t}-p_{s_l,t}, \quad \quad \quad  \forall l\in L, t=24,$ \\
   $x_{i,t} (1-x_{i-1,t}) =0,  \quad  \quad  \quad  \quad  \quad \forall i, i-1 \in I_{n,t}, t\in T, n\in N,$ &  (13) \\
   $p_{t,n} = P_{min} + \sum_{i\in I_{n,t}^s} (P^1_{i,t}-P^0_{i,t})x_{i,t},  \quad \quad \forall  t\in T , n\in N,$ & (14)\\
  $ x_{i,t} \geq 0, \quad  \quad \quad  \quad  \quad  \quad \quad  \quad \quad \quad  \quad  \quad  \quad   \forall i \in I_{n,t},  t\in T , n\in N,$  & (15) \\
   $ y_{b,t} \in \{0,1\} \quad  \quad  \quad \quad  \quad \quad \quad  \quad \quad  \quad   \quad \forall b \in B_n, n\in N, t\in T, $\\
   $f_{l,t} \in [ \underline{\tau}_{l,t},\overline{\tau}_{l,t}], \quad \overline{\mu}_{l,t},\underline{\mu}_{l,t},\overline{\rho}_{l,t}, \underline{\rho}_{l,t} \geq 0,   \quad  \quad \forall l\in L, t\in T,$ \\
   $ p_{n,t} \in [ P_{min}, P_{max}],  \quad \quad \quad \quad \quad \quad \quad \quad \quad \quad \quad \forall n\in N, t\in T.$
\end{tabular}
\end{table}

Constraint (1) defines supply-demand balance for each period in each zone. (2) ensures that accepted fraction of a segment does not exceed 1, non-convex bids are accepted at most one period by (3), and (4) satisfies the link relation of the non-convex bids. (5) satisfies the ramping limits on a line. (6)-(7) are the price-matching compatibility constraints and (8)-(12) show price-flow conditions. (13)-(14) ensure all segments with negative/positive surplus must be fully rejected/accepted. Bounds on the variables, non-negativity and integrality constraints are defined in (15).

The resulting model is an MIQCQP problem. Commercial solvers are not capable of solving our proposed model to optimality for real sized problems. In the literature, there are exact solution techniques like Bender's decomposition but they may not find a feasible solution within 10 minutes. Hence, a heuristic approach is needed to find at least one solution for practical reasons.

\section{Adaptive Tabu Search Algorithm}
Tabu search (TS), introduced by \cite{glover1986future}, is a neighborhood search algorithm to find a solution to an optimization problem. The main difference between TS and other neighborhood algorithms is the ability of escaping from the trap of local optimal solutions \cite{glover1989tabu,glover1990tabu}. 

There are different versions of TS that focus on intelligent escape and exploration techniques. Reactive TS prevents the cycle occurrence by automatically learning the optimal tabu list size \cite{battiti1994reactive,chiang1997reactive}. Parallel TS approaches aim to leverage the computational resources to solve large scale optimization problems \cite{crainic1997toward,talbi1998parallel,attanasio2004parallel}. \cite{nowicki2005advanced} adds big valley phenomenon with path-relinking technique to TS. \cite{zhang2007improved} introduces mutation operation of the genetic algorithm to the original TS. \cite{glover2006parametric} proposes a parametric branch and bound procedure based on TS instead of a tree search. While the prior studies are mainly deterministic, the literature also has probabilistic extensions of TS \cite{xu1996probabilistic,kochetov2001probabilistic,ghosh2003probabilistic}.


Adaptive tabu search (ATS) is a TS technique where diversification and intensification are in balance. There are two new features added to TS: adaptive search radius and back-tracking mechanism. Instead of regular search radius, \cite{sarawut2006adaptive} and \cite{puangdownreong2002system} discuss adaptive radius for a faster intensification and use back tracking for their diversification strategy. ATS is being used in a wide variety of areas such as assignment problems \cite{miao2014applying,xie2015evolving} and controller design  problems \cite{suyapan2017controller,ketthong2017design}.

\begin{figure}[!h]
	\includegraphics[scale = 0.3]{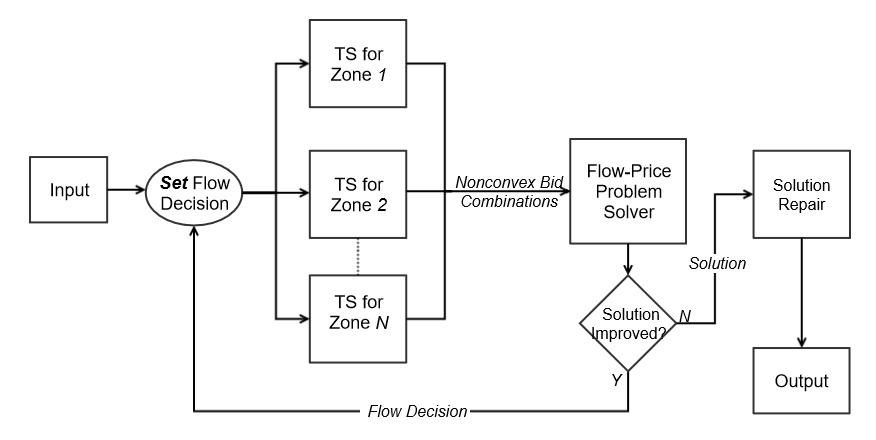}
   	\caption{Algorithm flow chart}
   		\label{flowchart}
\end{figure}

We apply a modified ATS to our problem because of the time restrictions in Turkish DAM. To solve \textit{Model-P}, we first relax the constraints defined in (8)-(12) so that each zone impacts another one through only flow variables ($f$). We assume that these flows are fixed to predetermined values by which we convert the problem into $|N|$ sub-problems. Each sub-problem represents different price regions. We solve these sub-problems by using TS with adaptive radius, and determine non-convex bid combinations for each zone. We use a commercial solver to find the optimal flows by fixing the integer variables to the values we found in TS. The resulting price-flow problem is a QP model defined by maximization of surplus subject to (1)-(5). Any optimal solution of the price-flow problem also satisfies (8)-(12) according to \cite{martin2014strict}. Then, we use the flows coming from the solver as a back-tracking and repeat the procedure until the objective function no longer improves. After the repetitions, we repair the solution to satisfy the requirements defined in (8)-(12). We propose a fast, multi-threaded algorithm by enduring the sub-optimality with this design. The flow chart of our proposed algorithm is presented in Figure \ref{flowchart}.

\subsection{Algorithm for Each Zone}
In this section, we give the details of the TS algorithm for solving the sub-problem of each zone. We start with the design decisions about tabu solution, tabu move, adaptive neighborhood, tabu list (TL), aspiration criteria, and  stopping condition.

We define \textit{tabu solution} for the sub-problems as the set of starting periods of the non-convex bids where 0 refers to rejection of the bid. We call \textit{tabu move} as changing one of the non-convex bid's starting period in a tabu solution. To illustrate, there is a tabu solution with two non-convex bids i.e., $S_a=\{b_1  \rightarrow  5,b_2 \rightarrow 0\}$. Tabu solution $S_a$ states that non-convex bid $b_1$ starts at period $5$, whereas $b_2$ is rejected. Then, by tabu move $b_1 \rightarrow0$, we create a new tabu solution $S_b=\{b_1 \rightarrow 0,b_2 \rightarrow 0\}$, where both of the non-convex bids are rejected.

By each tabu solution and fixed flow values, we can find the MCPs for the corresponding zone. A neighborhood around a tabu solution $S$ is represented as the rejection of accepted non-convex bid with positive surplus or acceptance of rejected non-convex bid with negative surplus with a random starting period. We also extend the neighborhood by randomly changing the starting period of an accepted non-convex bid. This size of neighborhood  decreases when we reach to local optima, because we expect that the number of rejected bids with positive surplus and accepted bids with negative surplus will be low. If a reject/accept move creates a tabu solution that violates constraint (4), we change the decision of corresponding child/mother of the bid until the resulting solution no longer violates (4). Hence, each neighbor in the neighborhood satisfies (4).

There may be some bids in neighbors where price-matching compatibility constraints are not satisfied because of the rejection of those bids. In such cases, we accept one of the violated non-convex bid and verify whether the compatibility constraints are satisfied. The procedure continues until there exists no bid in the neighbor that violates (6) or (7).


We keep a local TL and a global TL. The local TL keeps the tabu moves in a first-in-first-out queue. Global TL keeps the surplus of each tabu solution until the TS stops.  It prevents obtaining the same surplus in each iteration. The stopping condition in each zonal TS (\textit{Cond-1}) is limited by the number of iterations. 

We execute searches around both the region defined by constraints (1)-(4) and the region defined by (1)-(4) and (6)-(7). We take the solution with the best objective value. Each search consists of its own TL. Hence, a solution with a move that is in the TL of the other search can become a candidate solution for the next iteration. It creates the aspiration criteria. 

The initial solution $S_0$ is a tabu solution where all non-convex bids are accepted in a random period. After the stopping condition (\textit{Cond-1}) is reached, the algorithm jumps to the best solution found. The process repeats until the fourth jump, or the best solution is not improved.

The overall algorithm used for each region is shown in Table \ref{tablots}, where $g(S)$ represents the total surplus obtained from the tabu solution $S$.

\begin{table}[!h]
\caption{TS for each zone with adaptive radius}
\label{tablots}
\begin{tabular}{|l l|}
\hline
Step 1: & Start with a solution $S_0$. \\
Step 2: &If the stopping condition (Cond-1) does not hold. \\
		& Search around $S_0$: \\
		&	- By considering constraints (1)-(4) . Take the best solution $\overline{S}_0$ \\
        &   - By considering constraints (1)-(4), (6)-(7), (12)-(13). \\
        & Take the best solution $\underline{S}_0$ \\
Step 3:& Assign $S_0$ to  $\overline{S}_0$ if $g(S_0)> g(\overline{S}_0) $ or $\underline{S}_0$ for the reverse. \\ 
Step 4:& Update the global solution \\
        & Go Step 2. \\
\hline
\end{tabular}
\end{table}

\section{Numerical studies}
In this section, we explain how we generate the data we use in our experiments. Then, we present the experiment setup and results. 

\subsection{Data}
We create the data by considering possible market coupling scenarios of Turkey with its neighbor countries. Turkish electricity grid is connected to 8 countries. The maximum transmission capacity between Turkey and its neighbors is 1000 MW, but it may be restricted to 100 MW by the system operator to preserve the grid safety. 

\begin{figure}[!h]
\subfloat[]{
	\includegraphics[trim={3.5cm 25.4cm 14.8cm 1.4cm},clip,scale = 0.7]{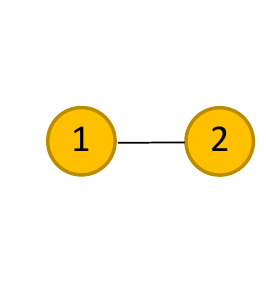}
}
\subfloat[]{
	\includegraphics[trim={6.5cm 25.4cm 11cm 1.4cm},clip,scale = 0.7]{TOPOLOGY.pdf}
}
\subfloat[]{
	\includegraphics[trim={10cm 25.4cm 7.4cm 1.4cm},clip,scale = 0.7]{TOPOLOGY.pdf}
}
    \caption{Zone topology for (a) 2, (b) 4 and (c) 8 zones}
    \label{topology}
\end{figure}

We constructed 9 data sets with 50 cases in each. Each set has different zone topology and flow scenarios. We assume that the zone topology does not change within the cases in each data set. Each zone topology consists of 2, 4 or 8 zones. Figure \ref{topology} shows the topology that has been used for the data creation. 
 
We assume that the bidders in each coupled zone have the similar bidding behaviors with the Turkish bidders. To create each case, we randomly select $|N|$ of the real Turkish DAM data between 1 June 2016 - 20 March 2018 by assuming each day represents the bids of a different zone. 

The ramping limits on the lines are ignored. The capacity and reverse capacity on the lines for each period are generated by using a uniform distribution between $0$ and $\alpha$. The value of $\alpha$ is equal to 0, 100, or 1000 MW in each data set.

Table \ref{bidinfo} presents the average number of bids per day when $\alpha$ is 0, 100, and 1000 MW and the number of zones are 2, 4 and 8. The first and the second number in the table indicate the average hourly and non-convex bids, respectively.

\begin{table}[!h]
\caption{The daily average number ($\#$) of hourly and non-convex bids for each data set}
\label{bidinfo}
\begin{tabular}{|c|c|c|c|}
\hline
\backslashbox{$\alpha$} {\# of zones} &2&4&8 \\ \hline
{0}   &31,245 - 311 & 62,373 - 618 & 124,725 - 1,217\\
\hline
{100} &31,286 - 307& 62,531 - 606 & 124,662 - 1,226\\
\hline
{1000} &31,192 - 306 & 62,354 - 623 & 124,431  - 1,216\\ 
\hline
\end{tabular}
\end{table}

\subsection{Experiment setup}
In our experiments, we apply our proposed ATS algorithm to data sets described in the previous section. We compare our algorithm with the exact and the heuristic solution methodologies proposed in \cite{martin2014strict}. We execute each method under 10 minutes in which EXIST must announce a market clearing solution. We compare the results in terms of time and surplus difference between the heuristic methods. We assume that each zone has the same bidding rules as in Turkish DAM.

The tests are performed on Intel Core i7-4790 CPU @ 3.60 GHz with 32 GB RAM configuration. We use IBM ILOG CPLEX 12.8 to solve the optimization problems.

To apply the solution methodologies proposed by \cite{martin2014strict} into our problem, we define a master problem as $\max g$ $ s.t.$ (1)-(4) and a sub-problem as finding a feasible solution to the constraints (7) - (15). Then, if the integer feasible solution in the branch-and-bound tree does not satisfy (7)-(15), we add a cut to the problem. The exact cut proposed by \cite{martin2014strict} also valid for our problem because it only cuts the current integer solution. However, the heuristic cut proposed by \cite{martin2014strict} allow the rejected bids with positive surplus, which violates the Turkish DAM rules. Hence, we modify the heuristic cut such that there should not be rejected bids with positive surplus. In our configuration, we define the heuristic cut as changing at least one reject decision of the bids which violate (6) or (7). We define this method as \textit{heuristic decomposition} in the rest of this study. 


%




\subsection{Results}
When we solve the model with the exact solution methodology proposed in \cite{martin2014strict}, we observe that there exist some cases without a feasible solution within 10 minutes. Hence, we focus on the results of ATS and heuristic decomposition. Table \ref{gapPerformance} and Table \ref{time} shows the surplus difference and time limits, respectively. 

\begin{table}[!h]
\caption{Average difference of the surplus values found by ATS and the heuristic decomposition}
\label{gapPerformance}
\centering
\begin{tabular}{|c|c|c|c|}
\hline
\backslashbox{$\alpha$} {\# of zones} &2&4&8 \\ \hline
{0}   & 1,911 & 6,763 & 10,482 \\
\hline
{100} &-3,686& -1,392 & 40,509\\
\hline
{1000} &-18,747 & -39,212 & -76,071 \\
\hline
\end{tabular}
\end{table}

Table \ref{gapPerformance} presents the surplus difference between ATS and the heuristic decomposition algorithm. Positive difference reflects that ATS performs better than the other one. When $\alpha=100$ and the number of zones is 4, there exists a case that the heuristic decomposition could not find a solution within the time limit. We compare the two algorithms for only the cases when there is at least one solution found. The results show that ATS provides better surplus values in the congested system whereas heuristic decomposition algorithm works better under the high transmission capacities. As the number of zone increases, parallel design structure of ATS gives an advantage to our approach.     

\begin{table}[!h]
\caption{Average run time (seconds) of ATS and the heuristic decomposition}
\label{time}
\centering
\begin{tabular}{|c|c|c|c|}
\hline
\backslashbox{$\alpha$} {\# of zones} &2&4&8 \\ \hline
{0}  & 8 - 106 & 16 - 513 & 37 - 600 \\
\hline
{100} & 10 - 107 &19 - 421 & 54 - 592\\
\hline
{1000} & 19 - 217  & 43 - 426 & 85 - 590 \\
\hline
\end{tabular}
\end{table}

Table \ref{time} shows the solution times of ATS and the heuristic decomposition method. ATS performs better in terms of solutions times in all of the configurations. When the number of zones increases, the average execution times of both algorithms increase. ATS is negatively affected by $\alpha$, i.e., when $\alpha$ increases ATS requires more time to converge to a solution. However, the heuristic decomposition is more robust for $\alpha$ changes. 

\section{Discussion and Conclusion}
In this study, we model the Turkish DAM clearing problem which consists of hourly, block and flexible bids, and network constraints. We propose an ATS algorithm to solve the resulting model. Since there is no real, zonal DAM data in EXIST, we generate random data sets under several assumptions. We show the performance of the algorithm by comparing its results with a heuristic decomposition method used in the literature, in terms of time and objective value.

The proposed ATS algorithm is a competitive alternative to the heuristic decomposition discussed in \cite{martin2014strict} since; (1) in some configurations the heuristic decomposition could not find any solution within the time limit, but ATS algorithm can find at least one feasible solution in all data sets, (2) ATS solves the problem faster for all data sets, and (3) ATS provides better surplus values than the heuristic decomposition under the low line capacities. 


\section*{Acknowledgment}
 The authors would like to thank The Scientific and Technological Research Council of Turkey – Technology and Innovation Funding Programs Directorate (T{\"U}B\.{I}TAK -TEYDEB) for supporting the project number 3161185. We also would like to thank Mustafa Kay{\i}r{\i}c{\i}, Birol Karatay, and Ozan G{\"u}rler for their supports on this work.



\bibliographystyle{IEEEtran}
\bibliography{bib}
%


\end{document}